\documentstyle[10pt,epsf,epsfig,hangcaption,xspace,amssymb,amsfonts,amsmath,amsthm,cite,
dp_delphititle]{dp_delphi}
\makeindex
\def\DpPaperGroup{EP}
\def\DpPaperRef{2001-062}
\def\DpDate{22 October 2001}
\def\DpAuthors{DELPHI Collaboration}
\def\DpSubmit{(Accepted by Phys. Lett. B)}
\def\DpTitle{{ Search for Charged Higgs Bosons in {\boldmath $e^+e^-$} 
Collisions at {\boldmath $\sqrt{s}$}=189--202~GeV}}
\def\DpComment{ }
\def\DpEMail{ }

\newcommand{\gev}{{\ifmmode \mbox{Ge\kern-0.2exV}
\else Ge\kern-0.2exV\nolinebreak\fi}}
\newcommand{\mev}{{\ifmmode \mbox{Me\kern-0.2exV}
\else Me\kern-0.2exV\nolinebreak\fi}}

\begin{document}
\makeatletter
\newcount\@tempcntc
\def\@citex[#1]#2{\if@filesw\immediate\write\@auxout{\string\citation{#2}}\fi
  \@tempcnta\z@\@tempcntb\m@ne\def\@citea{}\@cite{\@for\@citeb:=#2\do
    {\@ifundefined
       {b@\@citeb}{\@citeo\@tempcntb\m@ne\@citea\def\@citea{,}{\bf ?}\@warning
       {Citation `\@citeb' on page \thepage \space undefined}}%
    {\setbox\z@\hbox{\global\@tempcntc0\csname b@\@citeb\endcsname\relax}%
     \ifnum\@tempcntc=\z@ \@citeo\@tempcntb\m@ne
       \@citea\def\@citea{,}\hbox{\csname b@\@citeb\endcsname}%
     \else
      \advance\@tempcntb\@ne
      \ifnum\@tempcntb=\@tempcntc
      \else\advance\@tempcntb\m@ne\@citeo
      \@tempcnta\@tempcntc\@tempcntb\@tempcntc\fi\fi}}\@citeo}{#1}}
\def\@citeo{\ifnum\@tempcnta>\@tempcntb\else\@citea\def\@citea{,}%
  \ifnum\@tempcnta=\@tempcntb\the\@tempcnta\else
   {\advance\@tempcnta\@ne\ifnum\@tempcnta=\@tempcntb \else \def\@citea{--}\fi
    \advance\@tempcnta\m@ne\the\@tempcnta\@citea\the\@tempcntb}\fi\fi}
 
\makeatother
\begin{titlepage}
\pagenumbering{roman}
\CERNpreprint{\DpPaperGroup}{\DpPaperRef} 
\date{{\small\DpDate}} 
\title{\DpTitle} 
\address{\DpAuthors} 
\begin{shortabs} 
\noindent
%
\noindent

A search for pair-produced charged Higgs bosons was performed in the
high energy data collected by the DELPHI detector at LEP II at
centre-of-mass energies from 189~GeV to 202~GeV\@. The three different
final states, $\tau \nu \tau \nu$, $c \bar s \bar c s$ and $c \bar s
\tau \nu$ were considered. New methods were applied to reject
wrong hadronic jet pairings and for the tau identification, where a
discriminator based on tau polarisation and polar angles was used.  No
excess of data compared to the expected Standard Model processes was
observed and the existence of a charged Higgs boson with mass lower
than 71.5 GeV/$c^2$ is excluded at the 95\% confidence level.
\end{shortabs}
\vfill
\begin{center}
\DpSubmit \ \\ 
\DpComment \ \\
\DpEMail \ \\
\end{center}
\vfill
\clearpage
\headsep 10.0pt
\addtolength{\textheight}{10mm}
\addtolength{\footskip}{-5mm}
\begingroup
%
\newcommand{\DpName}[2]{\hbox{#1$^{\ref{#2}}$},\hfill}
\newcommand{\DpNameTwo}[3]{\hbox{#1$^{\ref{#2},\ref{#3}}$},\hfill}
\newcommand{\DpNameThree}[4]{\hbox{#1$^{\ref{#2},\ref{#3},\ref{#4}}$},\hfill}
\newskip\Bigfill \Bigfill = 0pt plus 1000fill
\newcommand{\DpNameLast}[2]{\hbox{#1$^{\ref{#2}}$}\hspace{\Bigfill}}
%
\footnotesize
\noindent
\DpName{J.Abdallah}{LPNHE}
\DpName{P.Abreu}{LIP}
\DpName{W.Adam}{VIENNA}
\DpName{P.Adzic}{DEMOKRITOS}
\DpName{T.Albrecht}{KARLSRUHE}
\DpName{T.Alderweireld}{AIM}
\DpName{R.Alemany-Fernandez}{CERN}
\DpName{T.Allmendinger}{KARLSRUHE}
\DpName{P.P.Allport}{LIVERPOOL}
\DpName{S.Almehed}{LUND}
% ...
\normalsize
\endgroup
\titlefoot{Department of Physics and Astronomy, Iowa State
     University, Ames IA 50011-3160, USA
    \label{AMES}}
\titlefoot{Physics Department, Universiteit Antwerpen,
     Universiteitsplein 1, B-2610 Antwerpen, Belgium \\
     \indent~~and IIHE, ULB-VUB,
     Pleinlaan 2, B-1050 Brussels, Belgium \\
     \indent~~and Facult\'e des Sciences,
     Univ. de l'Etat Mons, Av. Maistriau 19, B-7000 Mons, Belgium
    \label{AIM}}
\titlefoot{Physics Laboratory, University of Athens, Solonos Str.
     104, GR-10680 Athens, Greece
    \label{ATHENS}}
\titlefoot{Department of Physics, University of Bergen,
     All\'egaten 55, NO-5007 Bergen, Norway
    \label{BERGEN}}
\titlefoot{Dipartimento di Fisica, Universit\`a di Bologna and INFN,
     Via Irnerio 46, IT-40126 Bologna, Italy
    \label{BOLOGNA}}
\titlefoot{Centro Brasileiro de Pesquisas F\'{\i}sicas, rua Xavier Sigaud 150,
     BR-22290 Rio de Janeiro, Brazil \\
     \indent~~and Depto. de F\'{\i}sica, Pont. Univ. Cat\'olica,
     C.P. 38071 BR-22453 Rio de Janeiro, Brazil \\
     \indent~~and Inst. de F\'{\i}sica, Univ. Estadual do Rio de Janeiro,
     rua S\~{a}o Francisco Xavier 524, Rio de Janeiro, Brazil
    \label{BRASIL}}
\titlefoot{Coll\`ege de France, Lab. de Physique Corpusculaire, IN2P3-CNRS,
     FR-75231 Paris Cedex 05, France
    \label{CDF}}
\titlefoot{CERN, CH-1211 Geneva 23, Switzerland
    \label{CERN}}
\titlefoot{Institut de Recherches Subatomiques, IN2P3 - CNRS/ULP - BP20,
     FR-67037 Strasbourg Cedex, France
    \label{CRN}}
\titlefoot{Now at DESY-Zeuthen, Platanenallee 6, D-15735 Zeuthen, Germany
    \label{DESY}}
\titlefoot{Institute of Nuclear Physics, N.C.S.R. Demokritos,
     P.O. Box 60228, GR-15310 Athens, Greece
    \label{DEMOKRITOS}}
\titlefoot{Dipartimento di Fisica, Universit\`a di Genova and INFN,
     Via Dodecaneso 33, IT-16146 Genova, Italy
    \label{GENOVA}}
\titlefoot{Institut des Sciences Nucl\'eaires, IN2P3-CNRS, Universit\'e
     de Grenoble 1, FR-38026 Grenoble Cedex, France
    \label{GRENOBLE}}
\titlefoot{Helsinki Institute of Physics, HIP,
     P.O. Box 9, FI-00014 Helsinki, Finland
    \label{HELSINKI}}
\titlefoot{Joint Institute for Nuclear Research, Dubna, Head Post
     Office, P.O. Box 79, RU-101 000 Moscow, Russian Federation
    \label{JINR}}
\titlefoot{Institut f\"ur Experimentelle Kernphysik,
     Universit\"at Karlsruhe, Postfach 6980, DE-76128 Karlsruhe,
     Germany
    \label{KARLSRUHE}}
\titlefoot{Institute of Nuclear Physics,Ul. Kawiory 26a,
     PL-30055 Krakow, Poland
    \label{KRAKOW1}}
\titlefoot{Faculty of Physics and Nuclear Techniques, University of Mining
     and Metallurgy, PL-30055 Krakow, Poland
    \label{KRAKOW2}}
\titlefoot{Universit\'e de Paris-Sud, Lab. de l'Acc\'el\'erateur
     Lin\'eaire, IN2P3-CNRS, B\^{a}t. 200, FR-91405 Orsay Cedex, France
    \label{LAL}}
\titlefoot{School of Physics and Chemistry, University of Landcaster,
     Lancaster LA1 4YB, UK
    \label{LANCASTER}}
\titlefoot{LIP, IST, FCUL - Av. Elias Garcia, 14-$1^{o}$,
     PT-1000 Lisboa Codex, Portugal
    \label{LIP}}
\titlefoot{Department of Physics, University of Liverpool, P.O.
     Box 147, Liverpool L69 3BX, UK
    \label{LIVERPOOL}}
\titlefoot{LPNHE, IN2P3-CNRS, Univ.~Paris VI et VII, Tour 33 (RdC),
     4 place Jussieu, FR-75252 Paris Cedex 05, France
    \label{LPNHE}}
\titlefoot{Department of Physics, University of Lund,
     S\"olvegatan 14, SE-223 63 Lund, Sweden
    \label{LUND}}
\titlefoot{Universit\'e Claude Bernard de Lyon, IPNL, IN2P3-CNRS,
     FR-69622 Villeurbanne Cedex, France
    \label{LYON}}
\titlefoot{Dipartimento di Fisica, Universit\`a di Milano and INFN-MILANO,
     Via Celoria 16, IT-20133 Milan, Italy
    \label{MILANO}}
\titlefoot{Dipartimento di Fisica, Univ. di Milano-Bicocca and
     INFN-MILANO, Piazza della Scienza 2, IT-20126 Milan, Italy
    \label{MILANO2}}
\titlefoot{IPNP of MFF, Charles Univ., Areal MFF,
     V Holesovickach 2, CZ-180 00, Praha 8, Czech Republic
    \label{NC}}
\titlefoot{NIKHEF, Postbus 41882, NL-1009 DB
     Amsterdam, The Netherlands
    \label{NIKHEF}}
\titlefoot{National Technical University, Physics Department,
     Zografou Campus, GR-15773 Athens, Greece
    \label{NTU-ATHENS}}
\titlefoot{Physics Department, University of Oslo, Blindern,
     NO-0316 Oslo, Norway
    \label{OSLO}}
\titlefoot{Dpto. Fisica, Univ. Oviedo, Avda. Calvo Sotelo
     s/n, ES-33007 Oviedo, Spain
    \label{OVIEDO}}
\titlefoot{Department of Physics, University of Oxford,
     Keble Road, Oxford OX1 3RH, UK
    \label{OXFORD}}
\titlefoot{Dipartimento di Fisica, Universit\`a di Padova and
     INFN, Via Marzolo 8, IT-35131 Padua, Italy
    \label{PADOVA}}
\titlefoot{Rutherford Appleton Laboratory, Chilton, Didcot
     OX11 OQX, UK
    \label{RAL}}
\titlefoot{Dipartimento di Fisica, Universit\`a di Roma II and
     INFN, Tor Vergata, IT-00173 Rome, Italy
    \label{ROMA2}}
\titlefoot{Dipartimento di Fisica, Universit\`a di Roma III and
     INFN, Via della Vasca Navale 84, IT-00146 Rome, Italy
    \label{ROMA3}}
\titlefoot{DAPNIA/Service de Physique des Particules,
     CEA-Saclay, FR-91191 Gif-sur-Yvette Cedex, France
    \label{SACLAY}}
\titlefoot{Instituto de Fisica de Cantabria (CSIC-UC), Avda.
     los Castros s/n, ES-39006 Santander, Spain
    \label{SANTANDER}}
\titlefoot{Inst. for High Energy Physics, Serpukov
     P.O. Box 35, Protvino, (Moscow Region), Russian Federation
    \label{SERPUKHOV}}
\titlefoot{J. Stefan Institute, Jamova 39, SI-1000 Ljubljana, Slovenia
     and Laboratory for Astroparticle Physics,\\
     \indent~~Nova Gorica Polytechnic, Kostanjeviska 16a, SI-5000 Nova Gorica, Slovenia, \\
     \indent~~and Department of Physics, University of Ljubljana,
     SI-1000 Ljubljana, Slovenia
    \label{SLOVENIJA}}
\titlefoot{Fysikum, Stockholm University,
     Box 6730, SE-113 85 Stockholm, Sweden
    \label{STOCKHOLM}}
\titlefoot{Dipartimento di Fisica Sperimentale, Universit\`a di
     Torino and INFN, Via P. Giuria 1, IT-10125 Turin, Italy
    \label{TORINO}}
\titlefoot{Dipartimento di Fisica, Universit\`a di Trieste and
     INFN, Via A. Valerio 2, IT-34127 Trieste, Italy \\
     \indent~~and Istituto di Fisica, Universit\`a di Udine,
     IT-33100 Udine, Italy
    \label{TU}}
\titlefoot{Univ. Federal do Rio de Janeiro, C.P. 68528
     Cidade Univ., Ilha do Fund\~ao
     BR-21945-970 Rio de Janeiro, Brazil
    \label{UFRJ}}
\titlefoot{Department of Radiation Sciences, University of
     Uppsala, P.O. Box 535, SE-751 21 Uppsala, Sweden
    \label{UPPSALA}}
\titlefoot{IFIC, Valencia-CSIC, and D.F.A.M.N., U. de Valencia,
     Avda. Dr. Moliner 50, ES-46100 Burjassot (Valencia), Spain
    \label{VALENCIA}}
\titlefoot{Institut f\"ur Hochenergiephysik, \"Osterr. Akad.
     d. Wissensch., Nikolsdorfergasse 18, AT-1050 Vienna, Austria
    \label{VIENNA}}
\titlefoot{Inst. Nuclear Studies and University of Warsaw, Ul.
     Hoza 69, PL-00681 Warsaw, Poland
    \label{WARSZAWA}}
\titlefoot{Fachbereich Physik, University of Wuppertal, Postfach
     100 127, DE-42097 Wuppertal, Germany
    \label{WUPPERTAL}}
\addtolength{\textheight}{-10mm}
\addtolength{\footskip}{5mm}
\clearpage
\headsep 30.0pt
\end{titlepage}
%
\pagenumbering{arabic} 
\setcounter{footnote}{0} %
\large

\section{Introduction}

The existence of a charged Higgs boson doublet is predicted by several
extensions of the Standard Model. Pair-production of charged Higgs
bosons occurs mainly via $s$-channel exchange of a photon or a Z$^0$
boson. In two-doublet models, the couplings are completely specified
in terms of the electric charge and the weak mixing angle, $\theta_W$,
and therefore the production cross-section depends only on the charged
Higgs boson mass. Higgs bosons couple to mass and therefore decay
preferentially to heavy particles, but the details are model
dependent. We assume that at LEP energies $\tau \nu_\tau$ pair and a
$cs$ quark pair channels saturate the charged Higgs boson decays, and
analyses of the three possible final states, $\tau \nu \tau \nu$, $c
\bar s \bar c s$ and $c \bar s \tau \nu$, have been performed and are
described in this paper. The Higgs decay branching fraction to leptons
has been treated as a free parameter in the combination of the results
of these three analyses.

A search for pair-produced charged Higgs bosons was performed in
the data collected by DELPHI during the LEP runs at centre-of-mass
energies from 189~GeV to 202~GeV\@. The results reported here update
those obtained in an earlier analysis of the DELPHI data limited to
the 183~GeV run~\cite{delphihh}.  Similar searches have been
performed by the other LEP experiments~\cite{other_lep}.

A new technique was developed to improve the discrimination against
the hadronic W decays in the search for H$^{\pm}$ candidates.
Improved methods using the $\tau$ polarisation and boson production
angles in the leptonic and semileptonic channels were used for
rejection of W$^+$W$^-$ background.

\section{Data Analysis}

Data collected during the years 1998 and 1999 at centre-of-mass
energies from 189 GeV to 202 GeV were used. The total integrated
luminosity of these data samples is approximately 380 pb$^{-1}$.  The
DELPHI detector and its performance have already been described in
detail elsewhere~\cite{delphidet1,delphidet}\footnote{The co-ordinate
  system used has the z-axis parallel to the electron beam, and the
  polar angle calculated with respect to this axis.}.

Signal samples were simulated using the HZHA generator~\cite{HZHA}.
The background estimates from the different Standard Model processes
were based on the following event generators: PYTHIA~\cite{pythia} for
$q\bar{q}(\gamma)$, KORALZ~\cite{koralz} for $\mu^+\mu^-$ and
$\tau^+\tau^-$, BABAMC~\cite{bafo} for $e^+e^-$ and
EXCALIBUR~\cite{excalibur} for four-fermion final states. Two-photon
interactions were generated with TWOGAM~\cite{twogam} for hadronic
final states, BDK~\cite{bdk} for electron final states and
BDKRC~\cite{bdk} for other leptonic final states.

In all three analyses the final background rejection was performed by
using a likelihood technique. For each of the $N$ discriminating
variables, the fractions $F_{i}^{HH}(x_i)$ and $F_{i}^{bkg}(x_i)$ of
respectively H$^+$H$^-$ and background events, corresponding to a
given value $x_i$ of the $i^{th}$ variable, were extracted from 
samples of simulated H$^+$H$^-$ and background events normalised to
equal size. The signal likelihood was computed as the normalised
product of these individual fractions, $\prod_{i=1,N} F_{i}^{HH}(x_i)/
(\prod_{i=1,N} F_{i}^{HH}(x_i) + \prod_{i=1,N} F_{i}^{bkg}(x_i))$.

\subsection{The leptonic channel}

The signature for H$^+$H$^- \rightarrow \tau^+ \nu_\tau \tau^-
\bar{\nu}_{\tau}$ is large missing energy and momentum and two
acollinear and acoplanar \footnote{The acoplanarity is defined as the
  complement of the angle between the two jets projected onto the
  plane perpendicular to the beam.} jets containing either a lepton or
one or a few hadrons. Tight requirements for efficient operation of
the most important sub-detectors were used in this analysis in order
to ensure good quality of the tracks. These result in slightly smaller
integrated luminosities than in the hadronic channel (see
Table~\ref{tab:tabeff}).

\subsubsection{Event preselection}

To select leptonic events a total charged particle multiplicity
between 2 and 6 was required. All particles in the event were
clustered into jets using the LUCLUS algorithm~\cite{pythia}
($d_{join}=6.5$~GeV/$c$) and only events with two reconstructed jets
were retained. Both jets had to contain at least one charged particle
and at least one jet had to contain not more than one charged
particle. The angle between the two jets was required to be larger
than $30^\circ$.

Two-fermion and two-photon events were rejected by requiring an
acoplanarity larger than $13^\circ$ if both jets were in the barrel
region ($43^\circ<\theta<137^\circ$) and larger than $25^\circ$ otherwise.

The two-photon background was further reduced by the following energy
and momentum requirements: the sum of the jet energies multiplied by
the sines of their jet angles to the beam direction, $E_\perp$, was
required to be larger than $0.08 \sqrt{s}$ if both jets were in the
barrel region and larger than $0.1 \sqrt{s}$ in other cases; the total
transverse momentum, $p_\perp$, to be greater than $0.04 \sqrt{s}$;
the total energy detected within $30^\circ$ around the beam axis to be
less than $0.1 \sqrt{s}$; and the total energy outside this region to
be greater than $0.1 \sqrt{s}$.

Additional $\tau$ identification cuts were applied to reject WW events
where the W's have not decayed to $\tau\nu$. If the $\tau$ jet was
identified as an electron it had to have a momentum below
$0.13\sqrt{s}$ and an electromagnetic energy below $0.14\sqrt{s}$.
For muons the momentum had to be below $0.13\sqrt{s}$.  If a $\tau$
decay candidate particle was not identified as either a muon or an
electron, it was considered to be a hadron and accepted as a $\tau$
decay particle without further requirements.  Events in which the
invariant mass of either of the jets was more than 3~GeV/$c^2$ were
rejected.

The effects of the $\tau^+ \nu_\tau \tau^- \bar{\nu}_{\tau}$ selection
cuts are shown in Table~\ref{tab:tntnsel} for the combined
189--202~GeV sample.

\subsubsection{Final background discrimination}

After these selections most of the remaining background consists of
W$^+$W$^- \rightarrow \tau^+ \nu_\tau \tau^- \bar{\nu}_\tau$ events.
Events from both the H$^+$H$^-$ signal and the W$^+$W$^-$ background
have similar topologies and due to the presence of missing neutrinos
in the decay of each of the bosons, it is not possible to reconstruct
the boson mass.  There are two important differences, however, that
were used in order to discriminate the signal from the W$^+$W$^-$
background: the boson polar angle and the $\tau$ polarisation.

Assuming that the $\nu_\tau$ has a definite helicity, the polarisation
($P_\tau$) of tau leptons originating from heavy boson decays is
determined entirely by the properties of weak interactions and the
nature of the parent boson. The helicity configuration for the signal
is H$^- \rightarrow \tau^-_R \mbox{$\bar\nu_\tau$}^{\phantom +}_R$
(H$^+ \rightarrow \tau^+_L {\nu_\tau}^{\phantom +}_L$) and for the
W$^{\pm}$ boson background it is W$^- \rightarrow \tau^-_L
\mbox{$\bar\nu_\tau$}^{\phantom +}_R$ (W$^+ \rightarrow \tau^+_R
{\nu_\tau}^{\phantom +}_L$) resulting in $P_\tau^{\rm H}=+1$ and
$P_\tau^{\rm W}=-1$. The angular and momentum distributions depend on
polarisation and it is possible to build estimators of the $\tau$
polarisation to discriminate between the two contributions.

The $\tau$ decays were classified into the following categories: $e$,
$\mu$, $\pi$, $\pi+n\gamma$, $3\pi$ and others. The information on the
$\tau$ polarisation was extracted from the observed kinematic
distributions of the $\tau$ decay products, i.e.\ their angles and
momenta.  These estimators are equivalent to those used at LEP
I~\cite{ptauz}.  For charged Higgs boson masses close to the
threshold, the boost of the bosons is relatively small and the $\tau$
energies are similar to those of the $\tau$'s from Z$^0$ decays
(40--50~GeV).

\begin{figure}[ht!]
\begin{center}
\epsfig{file=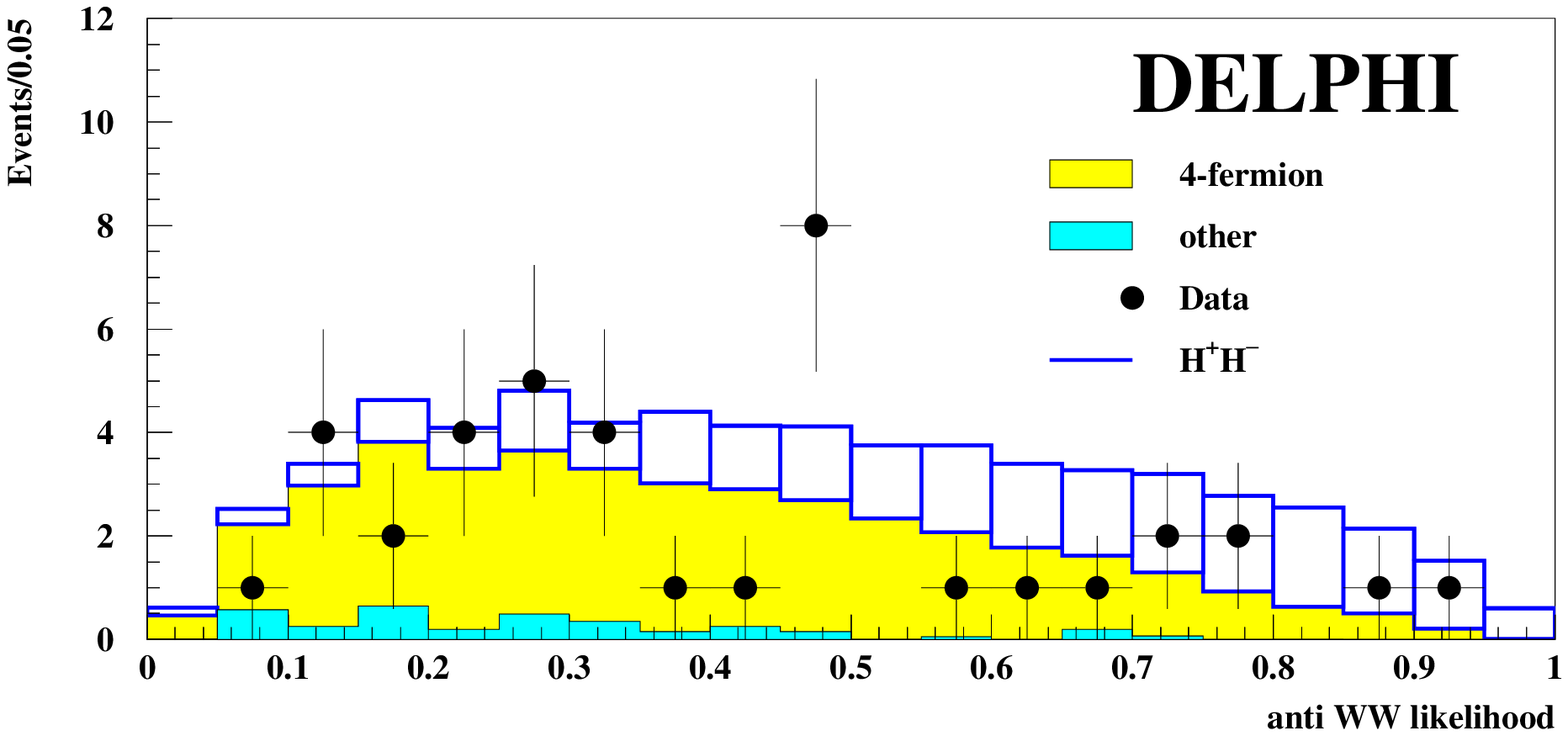,width=15.0cm}
\end{center}
\caption[]{Distribution of the anti-WW likelihood
  for leptonic events at 189--202~GeV\@. The expected histogram for a
  75 GeV/$c^2$ charged Higgs boson signal has been normalised to the
  production cross-section and 100\% leptonic branching ratio and
  added to the backgrounds.}

\label{fig:lik80}
\end{figure}

A likelihood to separate the signal from the W$^+$W$^-$ background was
built using four variables: the estimators of the $\tau$ polarisation
and the polar angle of the decay products of both $\tau$'s. The
distribution of that likelihood for data, expected backgrounds and
a 75~GeV/$c^2$ charged Higgs boson is shown in Fig~\ref{fig:lik80}.

\begin{table} [h!]
\begin{center}
\vspace{1ex}
\begin{tabular}{lrrrrr}
\hline
cut & \multicolumn{1}{c}{data}
    & \multicolumn{1}{c}{total bkg.}
    & \multicolumn{1}{c}{4-fermion}
    & \multicolumn{1}{c}{other bkg.}
    & \multicolumn{1}{c}{$\varepsilon_{75}$} \\ 
\hline

Leptonic selection  &  106274 & 108275    &  591 & 107684    &   73.6\% \\
Acoplanarity cut    &   10841 &  10519    &  452 &  10068    &   60.0\% \\
Energy/momentum cuts &       360 &    359    &  343 &     16    &   45.6\% \\
$\tau$ identification  &   39 &     39.7  &  36.3&      3.4  &   34.0\% \\
\hline

\end{tabular}
\end{center}
\caption{The total number of events observed and expected backgrounds
  in the leptonic channel after the different cuts used in the
  analysis. The last column shows the efficiency for a charged Higgs
  boson signal with $m_{\mathrm{H}^\pm} = 75$~GeV/$c^2$.}
\label{tab:tntnsel}
\end{table}

\subsection{The hadronic channel}

In the fully hadronic decay channel, each charged Higgs boson is
expected to decay into a $c \bar s$ pair, producing a four-jet final
state.  The two sources of background in this channel are the $q \bar
q g g$ QCD background and fully hadronic four-fermion final states.
In the four-fermion state background the significance of the
Z$^0$Z$^0$ pairs in the analysis is very small compared to the
W$^+$W$^-$ pairs. This is due to the lower cross-section for
Z$^0$Z$^0$ pair production and because the reconstructed Z$^0$Z$^0$
pair masses are concentrated at high masses, around the Z$^0$ boson
mass, out of the sensitivity reach of the analysis. Therefore, the
four-fermion sample is referred to as W$^+$W$^-$ in the rest of the
paper.

\subsubsection{Event preselection}

Events were clustered into four jets using the Durham
algorithm~\cite{durham}. The particle quality requirements and the
first level hadronic four-jet event selection followed in this
analysis were the same as for the DELPHI neutral Higgs
analysis~\cite{delphih0}.

In order to reject three-jet like QCD background events more
effectively, the Durham clustering parameter value for transition from
four to three jets ($y_{4 \rightarrow 3}$) was required to be greater
than 0.003. Events with a clear topology of more than four jets were
rejected by requiring the $y_{5 \rightarrow 4}$ value for transition
from five to four jets to be below 0.010 because of their worse di-jet
mass resolution after forcing them into four jets.

Energy-momentum conservation was imposed by performing a 4-C fit on
these events and the difference between the two di-jet masses for each
jet pairing was computed.  A 5-C fit, assuming equal boson masses, was
applied in order to improve the di-jet mass resolution. The di-jet
combination giving the smallest 5-C fit $\chi^2$ was selected for the
mass reconstruction. Events for which the 5-C fit $\chi^2$ divided by
the number of degrees of freedom exceeded 1.5 or the difference of the
masses computed with the same pairing after the 4-C fit exceeded
15~GeV/$c^2$ were rejected.

\subsubsection{Final background rejection}

The largest contribution to the part of the selected sample of
W$^+$W$^-$ events whose reconstructed mass is below the W mass peak
comes from picking one of the wrong di-jet pairings. These wrongly
paired events are characterised by a larger difference between the
masses of the two di-jets, i.e.\ the two boson candidates. As the
initial quark antiquark pairs are connected by a QCD colour field, in
which the hadrons are produced in the fragmentation process, the
wrongly paired events can also be identified using a method of colour
connection reconstruction~\cite{pt}.

The colour connection reconstruction method is based on the fact that,
in the rest frame of the correctly paired initial quark antiquark
pair, the hadrons that are produced in this colour string should have
small transverse momenta relative to the quark antiquark pair axis.
This could be distorted by hard gluon emission but such events are
suppressed with the $y_{5 \rightarrow 4}$ Durham parameter cut.  When
boosted into a rest frame of a wrongly paired quark pair the
transverse momenta of the particles relative to the quark quark axis
are larger.  The correct pairing is found by calculating the sum of
transverse particle momenta in each of the three possible pairing
hypotheses. The pairing chosen using the colour connection
reconstruction is compared to the pairing chosen using the
minimisation of the $\chi^2$ of the 5-C kinematical fit.  The output
of this comparison, called $p_\perp$-veto, is either agreement or
disagreement and it is used later in the analysis as one of the
variables in the background rejection likelihood.

The production polar angle of the positively charged boson
discriminates between W$^+$W$^-$ and Higgs pairs. This angle is
reconstructed as the polar angle of the di-jet with the higher sum of
jet charges, where the jet charge is calculated as a momentum weighted
sum of the charges of the particles in the jet~\cite{TGC}. The
distribution of this variable allows the discrimination of the signal
from the background of wrongly paired W$^+$W$^-$ events and QCD
events, even though in these latter cases the variable does not
correspond to a true boson production angle.

Since the charged Higgs boson is expected to decay to $c \bar s$ in
its hadronic decay mode, the QCD and W$^+$W$^-$ backgrounds can be
partially suppressed by selecting final states consistent with being
$c \bar s \bar c s$. A flavour tagging algorithm has been developed
for the study of multiparton final states \footnote{A similar jet
  flavour tagging technique has been used in a determination of
  $|V_{cs}|$ at LEP~II~\cite{jetag}.}. This tagging is based on nine
discriminating variables: three of them are related to the identified
lepton and hadron content of the jet, two depend on kinematical
variables and four on the reconstructed secondary decay structure. The
finite lifetime of $c$ (charm) particles is exploited to distinguish
between $c$ and light quark jets, while the $c$ mass and decay
multiplicity are used to discriminate against $b$ jets.  Furthermore
$s$ and $c$ jets can be distinguished from $u$ and $d$ jets by the
presence of an identified energetic kaon. Charged hadrons have been
identified using the combined response of RICH and TPC
$dE/dx$~\cite{part_id}. The responses of the flavour tagging algorithm
for the individual jets are further combined into an event $c \bar s
\bar c s$ probability.

The four variables described above: di-jet pair mass difference,the
$p_\perp$-veto, di-jet momentum polar angle and event $c \bar{s}
\bar{c} s$ probability, were combined to form an event anti-WW
likelihood function separating W$^+$W$^-$ events from H$^+$H$^-$
events. The response of this likelihood also discriminates H$^+$H$^-$
from the QCD background events.

An anti-QCD likelihood was formed using all four variables which were
used in the anti-WW likelihood and, in addition, two variables which
can be used to separate QCD background from pair-produced bosons: the
clustering algorithm parameter $y_{4 \rightarrow 3}$ and the event
acoplanarity.

\begin{figure}[ht!]
\begin{center}
  \epsfig{file=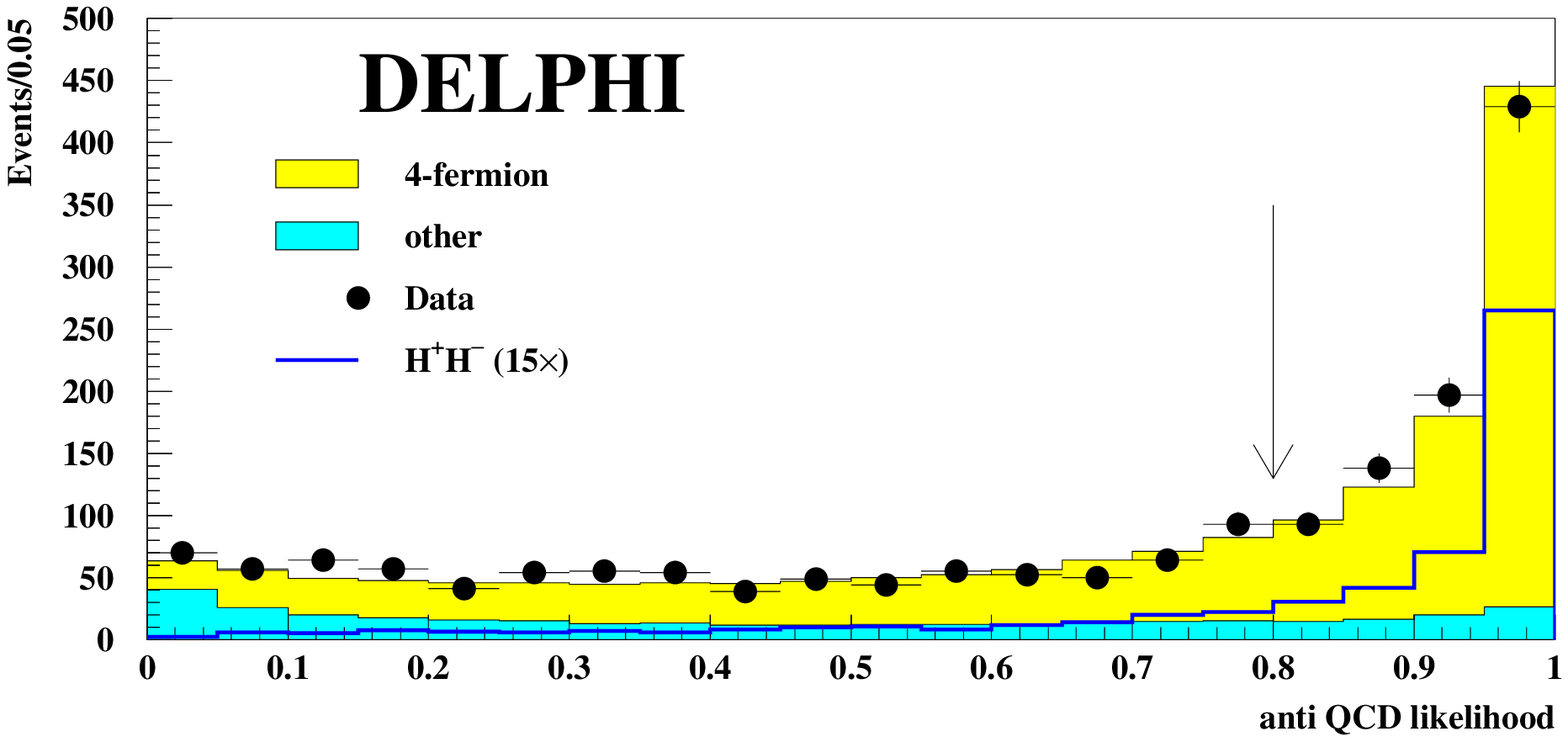,width=15.0cm}
  \epsfig{file=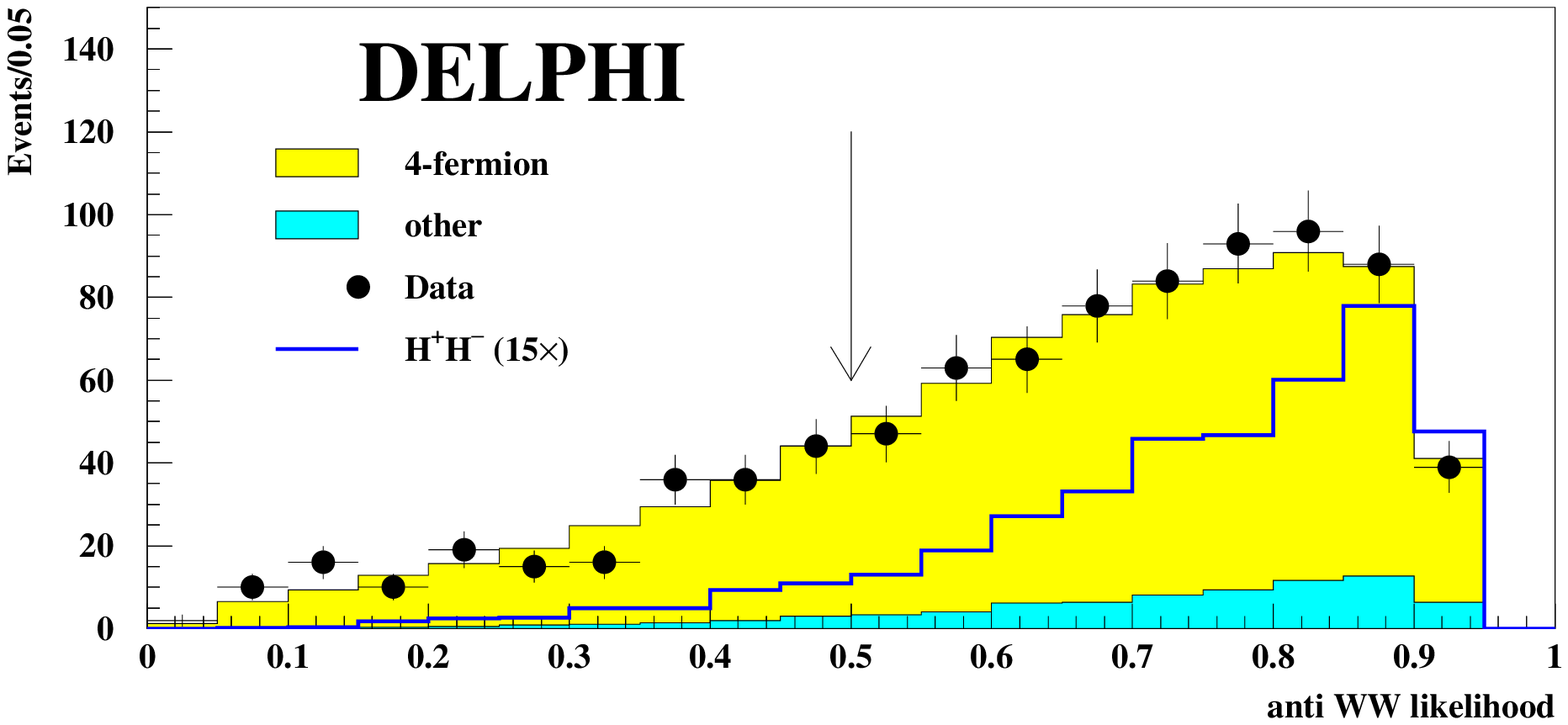,width=15.0cm}
\end{center}
\caption[]{Distributions of the anti-QCD and anti-WW likelihoods
  for hadronic events at 189--202~GeV\@. The anti-QCD likelihood is
  plotted on the preselection level and the anti-WW likelihood after a
  cut on the anti-QCD likelihood. The generated H$^+$H$^-$ signal mass
  is 75 GeV/$c^2$ and the signal histograms have been normalised to
  the production cross-section and 100\% hadronic branching ratio,
  multiplied by a factor of 15 and superimposed on the background
  histograms. The arrows indicate the cut values, below which events
  were rejected.}
\label{fig:cscslike}
\end{figure}

\begin{figure}[ht!]
\begin{center}
  \epsfig{file=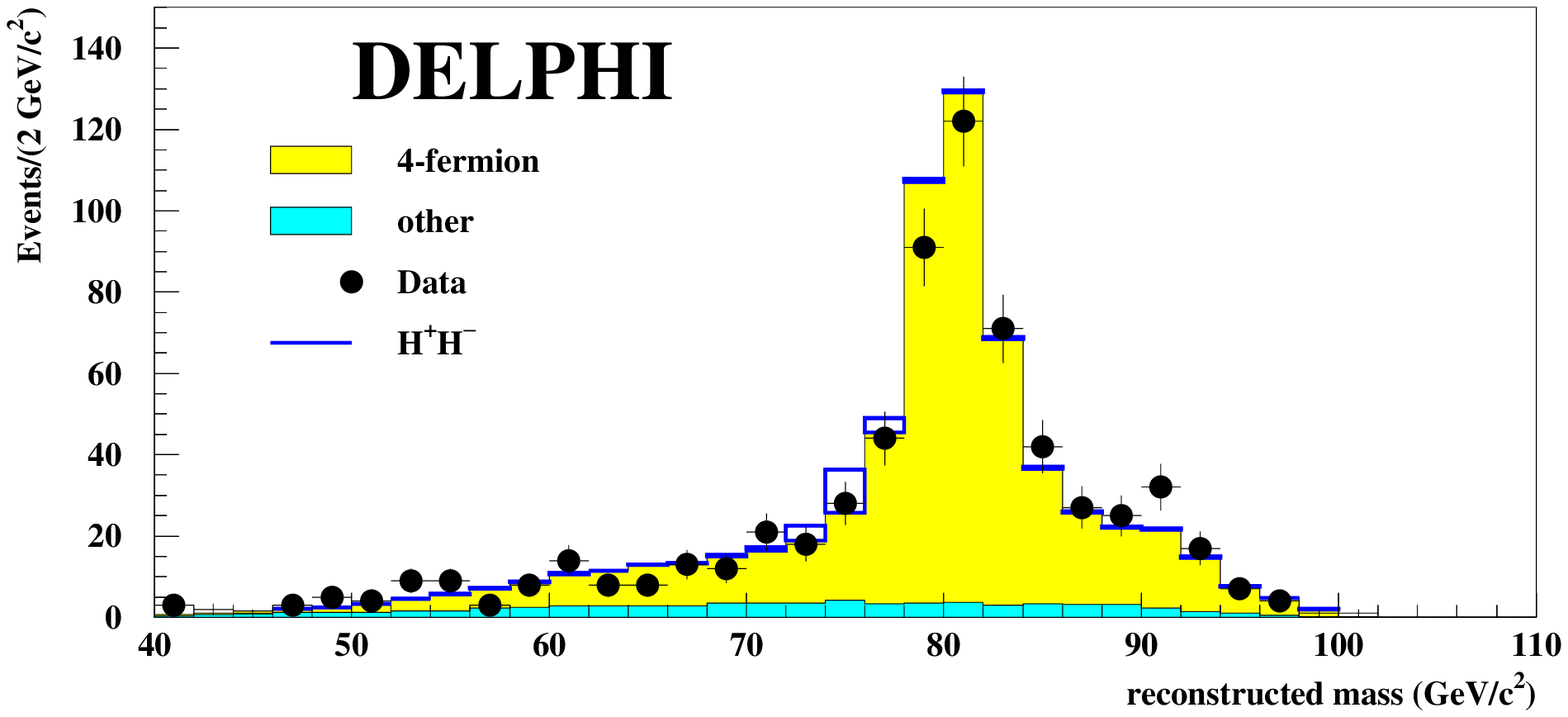,width=15.0cm}
\end{center}
\caption[]{Reconstructed mass distribution of hadronic events at 
  189--202~GeV at the final selection level. The generated
  H$^+$H$^-$ signal mass is 75 GeV/$c^2$ and the signal histogram
  has been normalised to the production cross-section and 100\%
  hadronic branching ratio and added to the backgrounds. }
\label{fig:cscsmass}
\end{figure}

The effects of the different sets of cuts are shown in
Table~\ref{tab:cscssel} for the combined 189--202~GeV sample. The
distribution of the anti-QCD likelihood on the preselection level and
the distribution of the anti-WW likelihood after a cut on the anti-QCD
likelihood are shown in Fig~\ref{fig:cscslike}. The reconstructed mass
distribution for data, expected backgrounds and signal after the 
anti-QCD and anti-WW cuts is shown in Fig~\ref{fig:cscsmass}.

\begin{table} [h!]
\begin{center}
\vspace{1ex}
\begin{tabular}{lrrrrr}

\hline
cut & \multicolumn{1}{c}{data}
    & \multicolumn{1}{c}{total bkg.}
    & \multicolumn{1}{c}{4-fermion}
    & \multicolumn{1}{c}{other bkg.}
    & \multicolumn{1}{c}{$\varepsilon_{75}$} \\ 
\hline
4-jet presel.      & 4156 &   3981.8 &   2515.0 &   1466.8 & 84.0\% \\
Durham $y_{cut}$   & 3066 &   2945.0 &   2110.0 &    835.0 & 71.4\% \\ 
$\chi^2$           & 2397 &   2340.0 &   1760.5 &    579.5 & 60.5\% \\
Mass diff.         & 1755 &   1708.6 &   1365.0 &    343.6 & 50.0\% \\
anti-QCD           &  857 &    844.5 &    767.0 &    77.5  & 36.5\% \\
anti-WW            &  653 &    645.7 &    577.8 &    67.9  & 33.2\% \\
\hline

\end{tabular}
\end{center}
\caption{The total number of events observed and expected backgrounds
  in the hadronic channel after the different cuts used in the
  analysis. The last column shows the efficiency for a charged Higgs
  boson signal with $m_{\mathrm{H}^\pm} = 75$~GeV/$c^2$.}
\label{tab:cscssel}
\end{table}

\subsection{The semileptonic channel}

In this channel one of the charged Higgs bosons decays into a $c \bar
s$ quark pair, while the other decays into $\tau\nu_\tau$. Such an
event is characterised by two hadronic jets, a $\tau$ candidate and
missing energy carried by the neutrinos. The dominating background
processes are QCD $q\bar{q}g$ event production and semileptonic decays
of W$^+$W$^-$. The same requirements for efficient operation of the most
important sub-detectors were used as in the analysis of the leptonic
channel.

\subsubsection{Event preselection and $\tau$ selection}

At least 15 particles, of which at least 8 were charged, were
required. The total energy of the observed particles had to exceed
0.30$\sqrt{s}$.  The missing transverse momentum had to be greater
than 0.08$\sqrt{s}$ and the modulus of the cosine of the angle between
the missing momentum and the beam had to be less than 0.8. Events were
also required to have no neutral particles with energy above 40 GeV\@.

After clustering into three jets using the Durham algorithm, the
clustering parameter $y_{3\rightarrow 2}$ was required to be greater
than 0.003, and each jet had to contain at least one charged particle.
The jet with the smallest charged particle multiplicity was treated as
the $\tau$ candidate and if two or more of the jets had the same
number of charged particles, the jet with smallest energy was chosen.
The $\tau$ candidate was required to have no more than six particles,
of which no more than three were charged.

\subsubsection{Final background rejection}

The mass of the decaying bosons was reconstructed using a constrained
fit requiring energy and momentum conservation with the known beam
energy and imposing the masses of the two bosons to be equal. The
three components of the momentum vector of the $\nu_\tau$ and the
magnitude of the $\tau$ momentum were treated as free parameters,
reducing the number of degrees of freedom of the fit from 5 to 1.
Only events with a reconstructed mass above 40 GeV/$c^2$ and a
$\chi^2$ below 2 were selected.

Separate likelihood functions were defined to distinguish the signal
events from the QCD and the W$^+$W$^-$ backgrounds, in a manner
similar to that used for the other channels described above.

To define the anti-QCD likelihood, the acollinearity of the event
(after forcing the event into two jets), the polar angle of the
missing momentum, the logarithm of the clustering parameter
$y_{3\rightarrow 2}$, and the product of the $\tau$ jet energy and the
smaller of the two angles between the $\tau$ jet and one of the other
jets were used as discriminating variables.

For the event anti-WW likelihood the variables used were the
reconstructed polar angle of the negatively charged boson (where the
charge was determined from the leading charged particle of the $\tau$
jet), the angle between the boson and the $\tau$ in the boson rest
frame, the energy of the $\tau$ jet, the classification of the decay
of the $\tau$ candidate ($e$, $\mu$, $\pi$, $\pi+n\gamma$, $3\pi$ and
others), and the $cs$ probability of the hadronic di-jet.

\begin{figure}[ht!]
\begin{center}
  \epsfig{file= 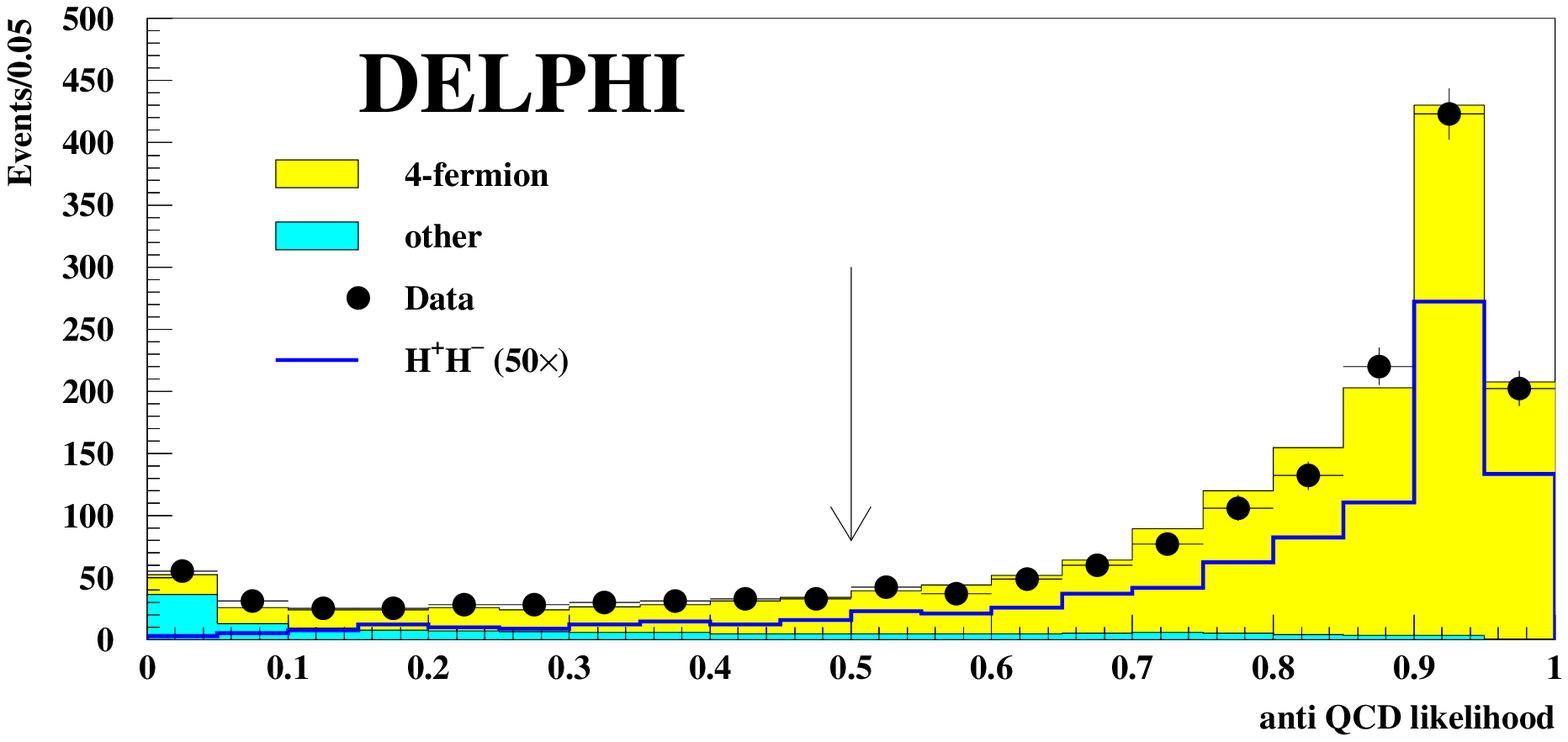,width=15.0cm}
  \epsfig{file= 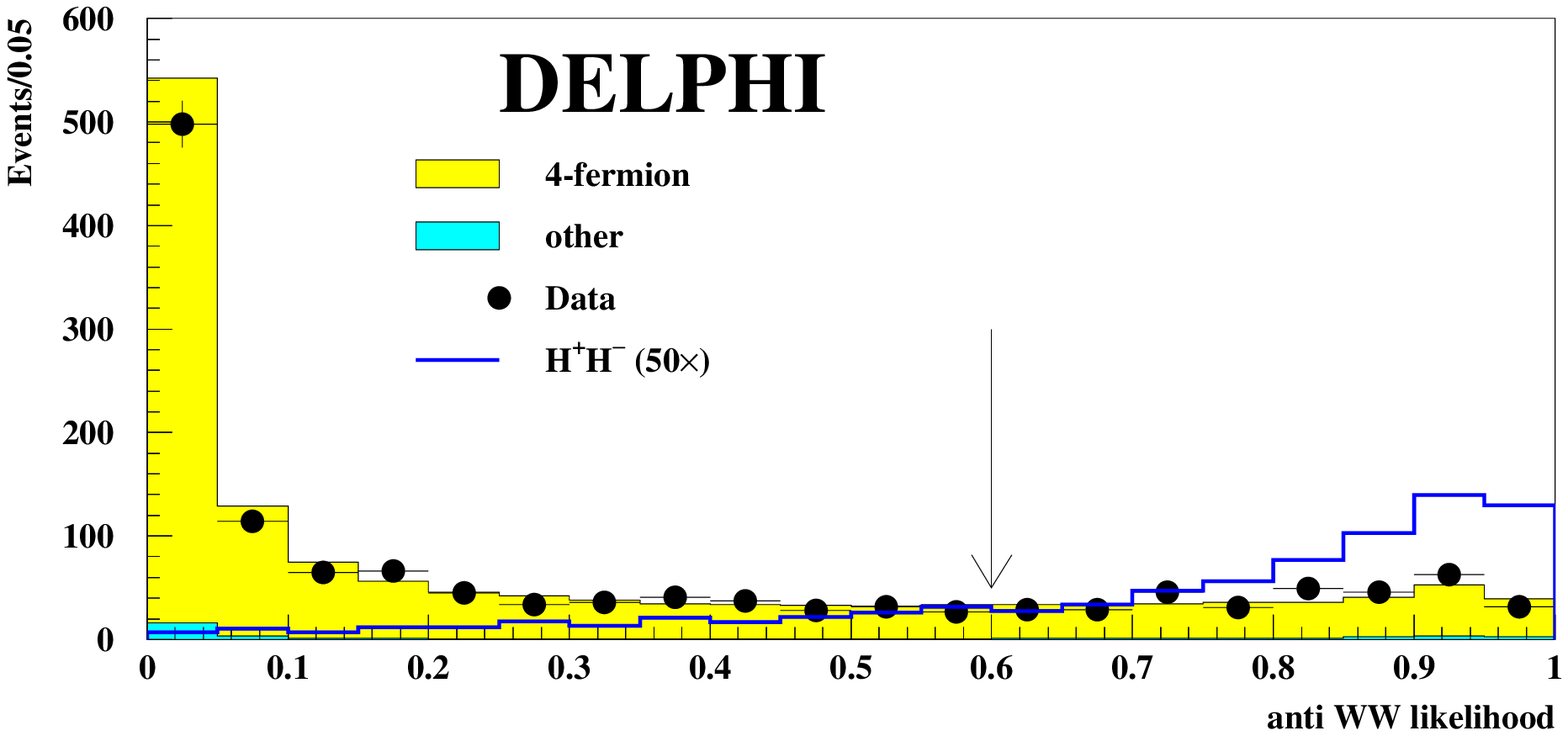,width=15.0cm}
\end{center}
\caption[]{Distributions of the anti-QCD and anti-WW likelihoods
  for semileptonic events at 189--202~GeV\@. The anti-QCD likelihood is
  plotted on the $\tau$ selection level and the anti-WW likelihood after a
  cut on the anti-QCD likelihood. The generated H$^+$H$^-$ signal mass
  is 75 GeV/$c^2$ and the signal histograms have been normalised to
  the production cross-section and 50\% leptonic branching ratio,
  multiplied by a factor of 50 and superimposed on the background
  histograms. The arrows indicate the cut values, below which events
  were rejected.}
\label{fig:cstnlike}
\end{figure}

\begin{figure}[ht!]
\begin{center}
  \epsfig{file= 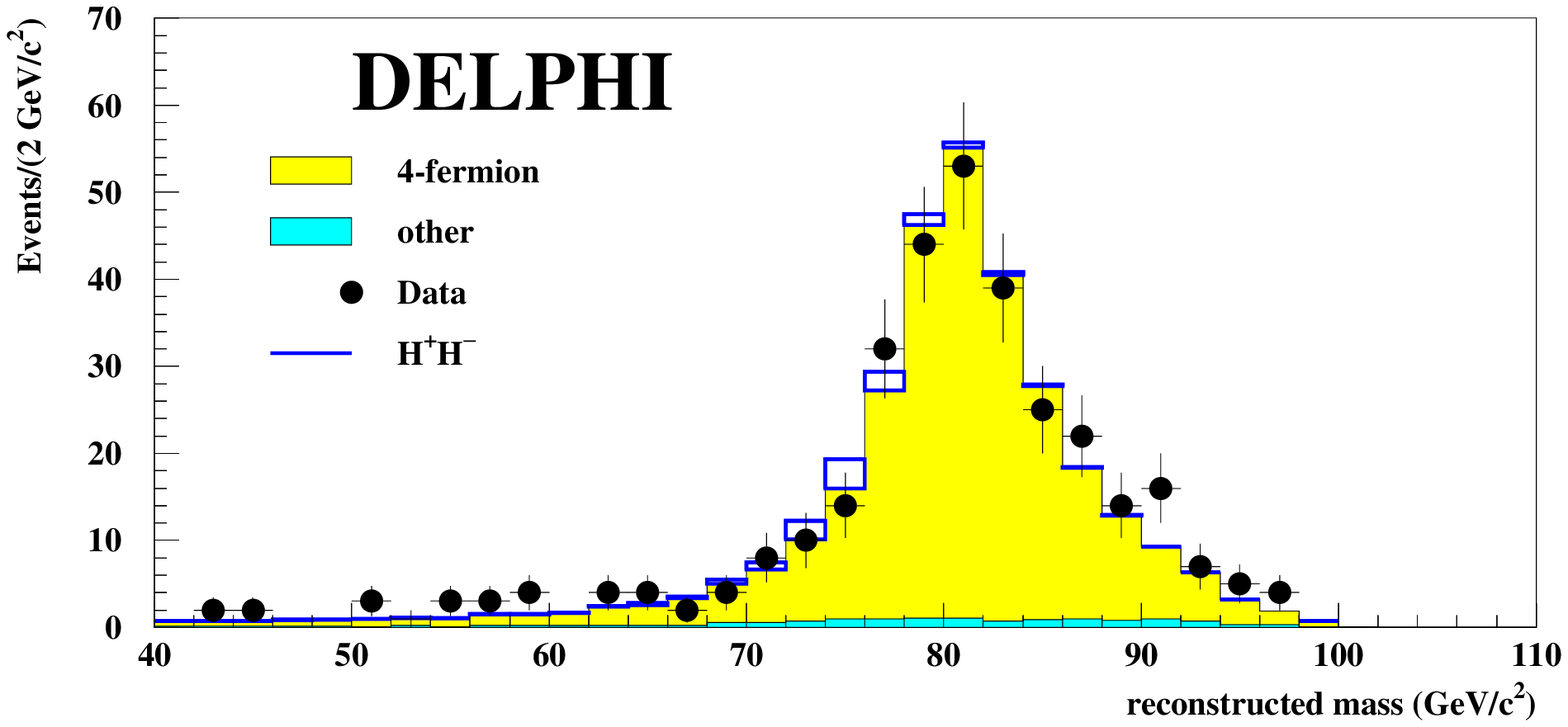,width=15.0cm}
\end{center}
\caption[]{Reconstructed mass distribution of semileptonic events at 
  189--202~GeV at the final selection level. The generated H$^+$H$^-$
  signal mass is 75 GeV/$c^2$ and the signal histogram has been
  normalised to the production cross-section and 50\% leptonic
  branching ratio and added to the backgrounds.}
\label{fig:cstnmass}
\end{figure}

The effects of the different sets of cuts are shown in
Table~\ref{tab:cstnsel} for the combined 189--202~GeV sample. The
distribution of the anti-QCD likelihood on the $\tau$ selection level
and the distribution of the anti-WW likelihood after a cut on the
anti-QCD likelihood are shown in Fig~\ref{fig:cstnlike}. The
reconstructed mass distribution for data, expected backgrounds and
signal after anti-QCD and anti-WW cuts is shown in
Fig~\ref{fig:cstnmass}.

\begin{table} [h!]
\begin{center}
\vspace{1ex}
\begin{tabular}{lrrrrr}

\hline
cut & \multicolumn{1}{c}{data}
    & \multicolumn{1}{c}{total bkg.}
    & \multicolumn{1}{c}{4-fermion}
    & \multicolumn{1}{c}{other bkg.}
    & \multicolumn{1}{c}{$\varepsilon_{75}$} \\ 
\hline
preselection     & 6395 & 6104.8 & 3043.8 & 3061.0 & 81.6\% \\
$\tau$ selection & 2149 & 2144.0 & 1788.0 &  356.0 & 58.1\% \\
$\chi^2$         & 1667 & 1699.0 & 1552.4 &  146.6 & 50.4\% \\
likelihoods      &  325 &  307.4 &  292.4 &   15.0 & 33.8\% \\
\hline

\end{tabular}
\end{center}
\caption{The total number of events observed and expected backgrounds 
  in the semi-leptonic channel after the different cuts used in the
  analysis. The last column shows the efficiency for a charged Higgs
  boson signal with $m_{\mathrm{H}^\pm} = 75$~GeV/$c^2$.}
\label{tab:cstnsel}
\end{table}

\section{Results}

\subsection{Selection efficiencies and uncertainties}

\begin{table}[h!]
\begin{center}
\begin{tabular}{cccrrr@{$\pm$}rr@{$\pm$}rr@{$\pm$}r}
\hline
& Chan. 
& $\sqrt{s}$ 
& \multicolumn{1}{c}{lum.}
& \multicolumn{1}{c}{data}
& \multicolumn{2}{c}{total bkg.}
& \multicolumn{2}{c}{$\varepsilon_{70}$} 
& \multicolumn{2}{c}{$\varepsilon_{75}$} \\
\hline 
\hline
& $\tau\nu\tau\nu$ & 189 & 153.8  & 16 & 15.0 & 1.5 & 32.3 & 1.6\% & 34.2 & 1.6\% \\
& $\tau\nu\tau\nu$ & 192 &  24.5  &  3 &  2.8 & 0.3 & 33.6 & 1.6\% & 34.2 & 1.6\% \\
& $\tau\nu\tau\nu$ & 196 &  72.4  & 10 &  8.6 & 0.8 & 33.6 & 1.6\% & 34.2 & 1.6\% \\
& $\tau\nu\tau\nu$ & 200 &  81.8  &  8 &  9.0 & 0.9 & 33.6 & 1.6\% & 33.5 & 1.6\% \\
& $\tau\nu\tau\nu$ & 202 &  39.4  &  2 &  4.4 & 0.4 & 33.6 & 1.6\% & 33.5 & 1.6\% \\
\hline
& $cscs$ & 189 &    154.3   &  288 & 267.8 & 16.1  & 36.2 & 2.0\% & 33.1 & 2.0\% \\
& $cscs$ & 192 &     25.5   &   36 &  42.7 &  2.6  & 36.2 & 2.0\% & 33.1 & 2.0\% \\  
& $cscs$ & 196 &     77.1   &  141 & 130.0 &  7.8  & 34.5 & 2.0\% & 33.8 & 2.0\% \\
& $cscs$ & 200 &     83.9   &  133 & 138.8 &  8.3  & 33.9 & 2.0\% & 33.5 & 2.0\% \\
& $cscs$ & 202 &     40.6   &   55 &  66.2 &  4.1  & 33.9 & 2.0\% & 33.5 & 2.0\% \\
\hline
& $cs\tau\nu$ & 189 & 153.8 & 126 & 118.8 & 6.9  & 33.0 & 1.8\% & 31.8 & 1.7\% \\
& $cs\tau\nu$ & 192 &  24.5 &  29 &   21.2 & 1.9 & 36.6 & 2.2\% & 35.2 & 2.1\% \\
& $cs\tau\nu$ & 196 &  72.4 &  76 &   62.2 & 5.5 & 36.6 & 2.2\% & 35.2 & 2.1\% \\ 
& $cs\tau\nu$ & 200 &  81.8 &  67 &   71.1 & 6.4 & 35.2 & 2.1\% & 35.3 & 2.1\% \\
& $cs\tau\nu$ & 202 &  39.4 &  27 &   34.1 & 3.1 & 35.2 & 2.1\% & 35.3 & 2.1\% \\
\hline
\end{tabular}
\end{center}
\caption[]{Integrated luminosity, observed number of events, 
expected number of background events and signal efficiency 
(70~GeV/$c^2$ and 75~GeV/$c^2$ masses) for 
different decay channels and centre-of-mass energies.} 

\label{tab:tabeff}

\end{table}

The number of real data and background events and the estimated
efficiencies for these selections for two different H$^{\pm}$ masses
are summarised in Table~\ref{tab:tabeff} for the three final states.
The quoted errors include the systematic uncertainties in the
expected background and the signal efficiency. Small contributions to
these uncertainties are due to uncertainties in the luminosity
measurement and in the cross-section estimates of the generated Monte
Carlo samples. 

The event selection and systematic errors in the leptonic analysis are
very similar to those in the DELPHI leptonic W$^+$W$^-$ analysis
\cite{ww189}. The largest part of the background and signal efficiency
uncertainties in the leptonic channel is due to the limited simulation
statistics available.  Combining these uncertainties gives a total
uncertainty of the order of 10\% in the background rate and 5\% in the
signal efficiency.

The largest contribution in the hadronic and semileptonic analyses is
due to differences in the distributions of the preselection and
likelihood variables in data and simulation. The systematic error on
the efficiency of the common DELPHI hadronic four-jet preselection has
been estimated to be $\pm4$\% \cite{delphih0}. The uncertainties
related to the other selection variables have been estimated by
comparing the shapes of the variable distributions in data and
simulation. This has been done at the preselection level where the
background event rate is so large that a possible signal would have no
effect on the global shapes of the variables. The agreement of all
variables has been found to be satisfactory to the level of a few
percent. Combining these errors, a total uncertainty of 6\% has
been estimated for the background rate and signal efficiency in the
hadronic channel. In the semileptonic channel the combined background
error estimate is 6 to 9\% depending on the energy sample
and the error of the signal efficiency is of the order of 6\%.
The combined error estimates are included in Table \ref{tab:tabeff}.

\subsection{Determination of the mass limit}
 
No significant signal-like excess of events was observed in any of the
three final states investigated. We find an agreement between data and
background expectations also in the mass region around 68 GeV/$c^2$
where the L3 collaboration has reported an excess using data collected
at the same centre-of-mass energies \cite{other_lep}.  A lower limit
for a charged Higgs boson mass was derived at 95\% confidence level as
a function of the leptonic Higgs decay branching ratio BR(H
$\rightarrow \tau\nu_\tau$).  The confidence in the signal hypothesis,
$CL_s$, was calculated using a likelihood ratio technique~\cite{alex}.

The background and signal probability density functions of one or two
discriminating variables in each channel were used. The data samples
collected at the five different centre-of-mass energies were treated
separately in the combination. Correlations between the systematic
uncertainties in different centre-of-mass energies were not included.
In the hadronic and semileptonic channels the two discriminating
variables were the reconstructed mass and the anti-WW likelihood; in
the leptonic channel only one background discrimination likelihood was
used since mass reconstruction is not possible. The distributions of
the discriminating variable for signal events, obtained by the
simulation at different H$^\pm$ mass values for each $\sqrt{s}$, were
interpolated for intermediate mass values.  To obtain the expected
signal rate at any given mass the signal efficiencies were fitted with
polynomial functions.

A Gaussian smearing of the central values of the number of expected
background events by their estimated uncertainties was introduced in
the limit derivation program.

\begin{figure}[ht!]
\begin{center}
\epsfig{file=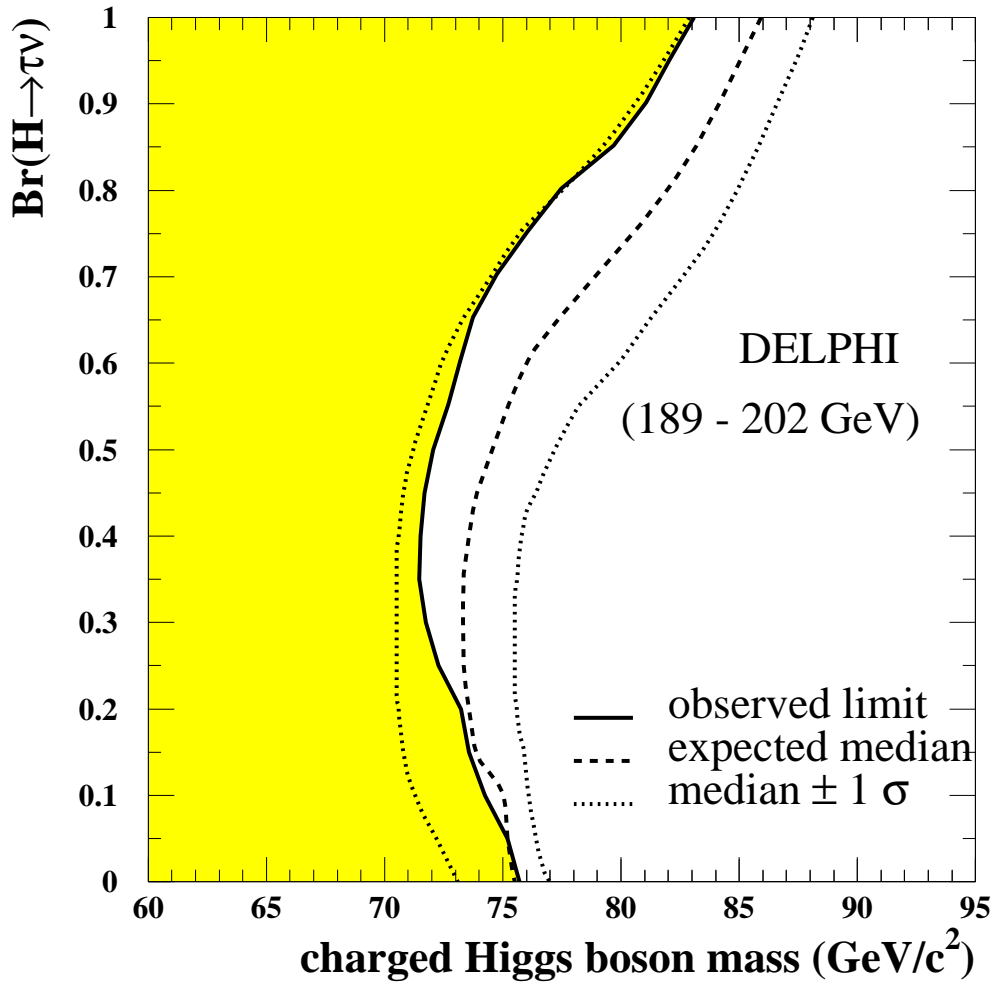,width=15.0cm}
\end{center}
\caption[]{The 95\% confidence level observed and expected exclusion 
  regions for H$^{\pm}$ in the plane BR($H \rightarrow \tau\nu_\tau$)
  vs.\ $M_{{\rm H}^{\pm}}$ obtained from a combination of the search
  results in the fully leptonic, hadronic and semileptonic decay
  channels at $\sqrt{s}=$ 189--202~GeV\@. The expected median of the
  lower mass limits has been obtained from a large number of simulated
  experiments. The median is the value which has 50\% of the limits of the
  simulated experiments below it and the $\pm$ 1$\sigma$ lines
  correspond similarly to 84\% and 16\% of the simulated experiments.}


\label{fig:limit}
\end{figure}

The results are summarised in Fig~\ref{fig:limit}. A lower H$^{\pm}$
mass limit of $M_{{\rm H}^{\pm}} >$ 71.5~GeV/$c^2$ can be set at the
95\% confidence level, independently of the branching ratio BR(H
$\rightarrow \tau\nu_\tau$). The production cross-section for the
charged Higgs bosons signal at this mass (71.5~GeV/$c^2$) and a
centre-of-mass energy of 202 GeV would be 0.255~pb. The median
of the limits obtained from a large number of simulated experiments is
73.3~GeV/$c^2$.

\section{Conclusion}

A search for pair-produced charged Higgs bosons was performed using
the full statistics collected by DELPHI at LEP at centre-of-mass
energies from 189~GeV to 202~GeV analysing the $\tau \nu \tau \nu$, $c
\bar s \bar c s$ and $c \bar s \tau \nu$ final states. No significant
excess of candidates was observed and a lower limit on the charged
Higgs mass of 71.5~GeV/$c^2$ is set at 95\% confidence level.  The
sensitivity of the DELPHI charged Higgs boson search is comparable
to the ones of the other LEP collaborations \cite{lep_hwg}.


\subsection*{Acknowledgements}
\vskip 3 mm
 We are greatly indebted to our technical 
collaborators, to the members of the CERN-SL Division for the excellent 
performance of the LEP collider, and to the funding agencies for their
support in building and operating the DELPHI detector.\\
We acknowledge in particular the support of \\
Austrian Federal Ministry of Science and Traffics, GZ 616.364/2-III/2a/98, \\
FNRS--FWO, Belgium,  \\
FINEP, CNPq, CAPES, FUJB and FAPERJ, Brazil, \\
Czech Ministry of Industry and Trade, GA CR 202/96/0450 and GA AVCR A1010521,\\
Danish Natural Research Council, \\
Commission of the European Communities (DG XII), \\
Direction des Sciences de la Mati$\grave{\mbox{\rm e}}$re, CEA, France, \\
Bundesministerium f$\ddot{\mbox{\rm u}}$r Bildung, Wissenschaft, Forschung 
und Technologie, Germany,\\
General Secretariat for Research and Technology, Greece, \\
National Science Foundation (NWO) and Foundation for Research on Matter (FOM),
The Netherlands, \\
Norwegian Research Council,  \\
State Committee for Scientific Research, Poland, 2P03B06015, 2P03B1116 and
SPUB/P03/178/98, \\
JNICT--Junta Nacional de Investiga\c{c}\~{a}o Cient\'{\i}fica 
e Tecnol$\acute{\mbox{\rm o}}$gica, Portugal, \\
Vedecka grantova agentura MS SR, Slovakia, Nr. 95/5195/134, \\
Ministry of Science and Technology of the Republic of Slovenia, \\
CICYT, Spain, AEN96--1661 and AEN96-1681,  \\
The Swedish Natural Science Research Council,      \\
Particle Physics and Astronomy Research Council, UK, \\
Department of Energy, USA, DE--FG02--94ER40817. \\

\end{document}